# Visualizing pair formation on the atomic scale in the high-$T_c$ superconductor $Bi_2Sr_2CaCu_2O_{8+\delta}$

Kenjiro K. Gomes[1*], Abhay N. Pasupathy[1*], Aakash Pushp[1*], Shimpei Ono[2], Yoichi Ando[2] & Ali Yazdani[1]

[1]Department of Physics, Joseph Henry Laboratories, Princeton University, Princeton, New Jersey 08544, USA. [2]Central Research Institute of Electric Power Industry, Komae, Tokyo 201-8511, Japan.

*These authors contributed equally to this work.

**Pairing of electrons in conventional superconductors occurs at the superconducting transition temperature $T_c$, creating an energy gap $\Delta$ in the electronic density of states (DOS)[1]. In the high-$T_c$ superconductors, a partial gap in the DOS exists for a range of temperatures above $T_c$ (ref. 2). A key question is whether the gap in the DOS above $T_c$ is associated with pairing, and what determines the temperature at which incoherent pairs form. Here we report the first spatially resolved measurements of gap formation in a high-$T_c$ superconductor, measured on $Bi_2Sr_2CaCu_2O_{8+\delta}$ samples with different $T_c$ values (hole concentration of 0.12 to 0.22) using scanning tunnelling microscopy. Over a wide range of doping from 0.16 to 0.22 we find that pairing gaps nucleate in nanoscale regions above $T_c$. These regions proliferate as the temperature is lowered, resulting in a spatial distribution of gap sizes in the superconducting state[3–5]. Despite the inhomogeneity, we find that every pairing gap develops locally at a temperature $T_p$, following the relation $2\Delta/k_BT_p = 7.9 \pm 0.5$. At very low doping ($\leq 0.14$), systematic changes in the DOS indicate the presence of another phenomenon[6–9], which is unrelated and perhaps competes with electron pairing. Our observation of nanometre-sized pairing regions provides the missing microscopic basis for understanding recent reports[10–13] of fluctuating superconducting response above $T_c$ in hole-doped high-$T_c$ copper oxide superconductors.**

The pseudogap state between $T_c$ and the temperature $T*$, below which the gap in the DOS occurs, has been the subject of a wide range of theoretical proposals—from those focused on superconducting pairing correlations without phase coherence[14,15] to those based on some form of competing electronic order or proximity to the Mott state[6–9]. Some experiments have shown evidence of pairing correlations above bulk $T_c$ (refs 10–13), while other experiments have associated the pseudogap with phenomena other than pairing, such as real space electronic organization, which is prominent at low dopings[16–18]. Various tunnelling experiments[19–21] have examined the relationship between the pseudogap and pair formation; however, the spatial variation of DOS and gaps on the nanoscale[3–5] complicates their interpretation and has prevented researchers from reaching a consistent conclusion. To obtain precise answers to these questions, it is necessary to perform systematic atomic-resolution scanning tunnelling microscopy (STM) measurements as a function of both doping and temperature.

We have measured the temperature evolution of the DOS of the high-$T_c$ superconductor $Bi_2Sr_2CaCu_2O_{8+\delta}$ using a specially designed variable temperature ultrahigh-vacuum STM. Our instrument allows us to track specific areas of the sample on the atomic scale as a function of temperature. To develop a systematic approach for the analysis of the data at different dopings and temperatures, we first consider the measurements on the most overdoped sample, with $x = 0.22$. At this doping, pseudogap effects have been reported to be either weak or absent[2,6], allowing us to interpret gaps in the DOS as those associated with superconductivity.

Figure 1 shows spectroscopy measurements for an overdoped sample with $T_c = 65$ K (OV65) performed at a specific atomic site over a range of temperatures close to $T_c$. From such spectra, we determine two important quantities. First, we measure the maximum value of the local gap $\Delta \approx 24$ meV. Second, we estimate the temperature $T_p \approx 72$–80 K at which $\Delta$ is no longer measurable, using the criterion that at this temperature $\mathrm{d}I/\mathrm{d}V (V = 0) \geq \mathrm{d}I/\mathrm{d}V$ (for all $V > 0$). Above $T_p$ the spectra shows a bias-asymmetric background in the DOS that changes little with increasing temperature—indicating that the pairing gap is either absent or no longer relevant at this atomic site. On the basis of this procedure, we find that the data in Fig. 1a can be described by the relation $2\Delta/k_B T_p \approx 7.7$, where $k_B$ is Boltzmann's constant and $T_p$ is the gap closing temperature. While this measurement at a single atomic site is not statistically significant, it establishes the procedure that we extend to large sets of similar measurements.

The evolution of the pairing gap with temperature can be examined statistically using spectroscopic mapping measurements over large areas (~300 Å) of the sample as a function of temperature. Such experiments allow direct visualization on the atomic scale of the development of gaps. In the superconducting state ($T < T_c$), the overdoped OV65 sample shows (Fig. 2a) a distribution of $\Delta$ (refs 3 and 4). With increasing temperature, gap maps and related histograms on the OV65 sample display a rapid increase of ungapped regions (Fig. 2a–e). Although the temperatures over which the gaps collapse locally are relatively close to $T_c$ for this sample, these measurements clearly demonstrate that gaps persist locally on the nanoscale over a range of temperatures $T_p$. These results are reminiscent of recent calculations of the evolution of pairing gaps with temperature in a $d$-wave superconductor with spatially varying pairing correlations[22].

A great deal of information about the nucleation of pairing on the atomic scale can be extracted from data in Fig. 2a–d. Here, we focus on extracting the relation between a given local $\Delta$ measured at $T < T_c$ and the temperature $T_p$ at which it collapses. From the gap maps in Fig. 2a–d, we can extract the percentage of the sample that is gapped at a given temperature (points in Fig. 2f). To compare, we use the histogram of $\Delta$ values measured at the lowest temperature to compute the probability $P(<\Delta)$ that the gaps are less than a given $\Delta$ (solid line). A linear relationship between local $\Delta$ and $T_p$ would require that the $x$ axis of these two measurements be related by a simple ratio. The best-fit ratio extracted in Fig. 2f is $2\Delta/k_B T_p = 7.8 \pm 0.3$. This relation shows that despite the strong variation of the superconducting gaps on the nanoscale in the overdoped sample, they all collapse following the same local criterion.

Having established the relation between local $\Delta$ and $T_p$ for the pairing gaps in the OV65 sample, we can study the temperature evolution of gaps in the DOS measured on samples with different dopings. In Fig. 3a–d, we show such measurements on an optimally doped sample OP93. Similar to overdoped samples, the low-temperature ($T < T_c$) spectra for the optimally doped sample are consistent with that of a $d$-wave pairing gap (see Supplementary Information). In contrast to the overdoped sample, which develops ungapped regions rapidly while crossing $T_c$, the optimal doped sample is still entirely gapped 10 K above its $T_c$. The loss of phase coherence at $T_c$ only affects the sharpness of the peaks in the spectra at $V = \pm\Delta$, while the gap in the superconducting state smoothly evolves into that measured above $T_c$ (see Fig. 3a)[19]. High-resolution gap maps measured at different temperatures (Fig. 3b–d) show that the distributions of gaps just above and below $T_c$ are essentially the same except for some broadening (see Supplementary Information). A further increase in the temperature results in an inhomogeneous collapse of gaps. The spatial collapse of the gaps is comparable to that observed in the OV65 sample (Fig. 2), except that the temperature range for $T_p$ values over which gaps collapse is much larger for the OP93 sample (105–160 K) than for the OV65 sample (64–80 K).

We can use the comparison between $P(<\Delta)$ measured at the lowest temperature with the percentage of the ungapped regions measured as a function of temperature to test our local pairing hypothesis for samples at various dopings. The measurements of these two quantities are displayed in Fig. 3e and f, where a single temperature–gap scaling relation $2\Delta/k_B T = 8.0$ has been used to plot data on all samples in this study. From Fig. 3e it is clear that overdoped and optimally doped samples have identical gap–temperature scaling ratios, which, together with the consistency of their low-temperature spectra with a $d$-wave superconducting gap, implies that we can interpret these gaps as those due to pairing. These results clearly show that pairing gaps and the temperature at which they collapse (which can be equal to or larger than $T_c$) follow a universal local criterion over a wide range of doping. The extracted ratio also shows that the local pairing gap is far more fragile to increases in temperature than are the conventional Bardeen–Cooper–Schrieffer (BCS) superconductors, for which the ratio is in the range of 3.5–5. Surprisingly, the ratio is insensitive to the size of the gap, for gaps ranging from ~15 to 50 mV, indicating that even the smallest gaps are very far from the BCS limit.

Our local pairing hypothesis, however, appears to fail in the underdoped regime (Fig. 3f). Although such a pairing hypothesis has its shortcomings, such as ignoring the possibility of a proximity effect, we show that the deviation from this picture in underdoped samples is most probably due to the complication of two energy scales in this doping regime. Fig. 4a shows representative spectra measured at temperatures far above $T_c$ that allow us to compare the behaviour of overdoped, optimal and underdoped samples. Once the pairing gaps collapse at high temperatures, the overdoped and optimally doped samples show remarkably similar electron–hole asymmetric spectra. In contrast, the underdoped UD73 sample shows very different, V-shaped spectra with an ill-defined gap, which is insensitive to increasing temperature. Clearly, such V-shaped spectra are related to the strong pseudogap behaviour in underdoped copper oxides[17,18]; however, these spectra and their pseudogap behaviour are qualitatively different from the pseudogaps observed on optimal and overdoped samples above $T_c$. Such a difference is also seen in ARPES (Angle-Resolved Photoelectron Spectroscopy) measurements where the gap closes in overdoped samples but 'fills in' for underdoped samples[23]. A key question is whether all gaps in the underdoped regime as measured by STM can be ascribed to pairing. We find that for $T \ll T_c$, more than 30% of the spectra on the underdoped sample show 'kinks' in the spectra at low bias, indicating the importance of a lower energy scale, as shown in Fig. 4b[3,17,18,24]. The probability of observing such spectra in optimal or overdoped samples is negligible. Spectroscopic mapping with STM can be used to determine the spatial variation and distribution of each energy scale (inset, Fig. 4b). The presence of this additional energy scale indicates that the large gaps seen in underdoped samples cannot be simply associated with pairing. Evidence for two energy scales, one related to pairing and one related to strong pseudogap behaviour in underdoped copper oxides, has been accumulating from photoemission[23–25] and Raman measurements[26]. In contrast, at doping levels beyond optimal doping, results from these experiments can be captured with a single energy scale[23,26]. Momentum resolution (lacking in STM) is important to resolve the two energy scales in underdoped samples. Further work is required to establish firmly whether the 'kink' energy in STM spectra is related to pairing in these samples.

Our ability to visualize the development of gaps and the local pairing hypothesis, which we have established quantitatively on optimal and overdoped samples provides a microscopic picture with which to understand several key aspects of the copper oxide phase diagram. In Fig. 5, we summarize our observations of the spatially inhomogeneous development of the gaps with a colour plot showing the percentage of areas that are gapped as a function of temperature at specific dopings. The rising percentage of the gapped regions in various samples with lowering temperature that we probed in our experiments on the atomic scale correlates remarkably well with the onset of the suppression of low-energy excitations probed by other techniques[27]. As we discussed above, for the optimal and overdoped samples a single energy gap can describe all of our findings (Figs 1–3 and additional details in the Supplementary Information) strongly suggesting that the onset of the gap is indeed due to pairing, which occurs locally at $T_p$. The apparent $T*$ line is controlled by the largest pairing gaps ($T_{p, max}$) for these samples.

In contrast, for underdoped samples our data supports the notion of two energy scales. The $T*$ line is controlled by the larger of the two scales, which appears to be unrelated to pairing. Our data suggests that the $T_{p, max}$ line could be well below $T*$ for low doping. This possibility is also supported by measurements of fluctuating superconductivity, which have shown that the onset temperatures for these fluctuations are well below $T*$ for underdoped samples[10–13]. A comparison of our data to those from Nernst and magnetization measurements on similar $Bi_2Sr_2CaCu_2O_{8+\delta}$ samples[13] shows that the macroscopic signature of the fluctuating superconducting state appears when ~50% of the sample develops a pairing gap. Given that an adequate amount of pairing has to develop in the samples for the vortex response to be well defined, our measurements provide the missing microscopic basis for the onset of the vortex response. Our observations of local pairing over a range of temperatures, as well as the Nernst and magnetization measurements, show that $T_c$ in the $Bi_2Sr_2CaCu_2O_{8+\delta}$ system marks the onset of phase coherence and not the formation of pairs as in BCS superconductors.

Finally, the local pairing criterion extracted from a large number of measurements on $Bi_2Sr_2CaCu_2O_{8+\delta}$ at various dopings and temperatures has important implications for the mechanism of pairing in copper oxides. In conventional superconductors, the strength of the coupling of electrons to phonons determines both $\Delta$ and $T_c$, with stronger coupling resulting in an increase of both these quantities. However, in the strong-coupling limit, the ratio $2\Delta/k_B T_c$ is dependent on $\Delta$ and increases from the universal BCS ratio of 3.5 (ref. 1). The extension of BCS theory based on the Eliashberg equations captures this behaviour for conventional superconductors[28]. Our observation of the insensitivity of the large ratio $2\Delta/k_BT_c = 7.9 \pm 0.5$ to the size of local $\Delta$ values (from 15–50 mV), the local disorder, as well as the doping is hence fundamentally different from the expectations[29] from a electron–boson pairing mechanism based on an Eliashberg-type theory. A successful theory of copper oxides would have to explain not only how pairing correlations can nucleate in isolated nanoscale regions at high temperatures but also the robustness of the local pairing criterion reported here.

1.  Tinkham, M. *Introduction to Superconductivity* (McGraw-Hill, New York, 1975).

2.  Timusk, T. & Statt, B. The pseudogap in high-temperature superconductors: an experimental survey. *Rep. Prog. Phys.* **62**, 61-122 (1999).

3.  Howald, C., Fournier, P. & Kapitulnik, A. Inherent inhomogeneities in tunnelling spectra of $Bi_2Sr_2CaCu_2O_{8-x}$ crystals in the superconducting state. *Phys. Rev. B* **64**, 100504(R) (2001).

4.  Pan, S. H. *et al.* Microscopic electronic inhomogeneity in the high-$T_c$ superconductor $Bi_2Sr_2CaCu_2O_{8+x}$. *Nature* **413**, 282-285 (2001).

5.  McElroy, K. *et al.* Atomic-scale sources and mechanism of nanoscale electronic disorder in $Bi_2Sr_2CaCu_2O_{8+\delta}$. *Science* **309**, 1048-1052 (2005).

6.  Tallon, J. L. & Loram, J. W. The doping dependence of T* - what is the real high-$T_c$ phase diagram? *Physica C* **349**, 53-68 (2001).

7.  Kivelson, S. A. *et al.* How to detect fluctuating stripes in the high-temperature superconductors. *Rev. Mod. Phys.* **75**, 1201-1241 (2003).

8.  Norman, M. R., Pines, D. & Kallin, C. The pseudogap: friend or foe of high $T_c$? *Adv. Phys.* **54**, 715-733 (2005).

9. Lee, P. A., Nagaosa, N. & Wen, X.-G. Doping a Mott insulator: Physics of high-temperature superconductivity. *Rev. Mod. Phys.* **78**, 17-86 (2006).

10. Corson, J., Mallozzi, R., Orenstein, J., Eckstein, J. N. & Bozovic, I. Vanishing of phase coherence in underdoped $Bi_2Sr_2CaCu_2O_{8+\delta}$. *Nature* **398**, 221-223 (1999).

11. Xu, Z. A., Ong, N. P., Wang, Y., Kakeshita, T. & Uchida, S. Vortex-like excitations and the onset of superconducting phase fluctuation in underdoped $La_{2-x}Sr_xCuO_4$. *Nature* **406**, 486-488 (2000).

12. Wang, Y. *et al.* Field-enhanced diamagnetism in the pseudogap state of the cuprate $Bi_2Sr_2CaCu_2O_{8+\delta}$ superconductor in an intense magnetic field. *Phys. Rev. Lett.* **95**, 247002 (2005).

13. Wang, Y., Li, L. & Ong, N. P. Nernst effect in high-$T_c$ superconductors. *Phys. Rev. B* **73**, 024510 (2006).

14. Emery, V. J. & Kivelson, S. A. Importance of phase fluctuations in superconductors with small superfluid density. *Nature* **374**, 434-437 (1995).

15. Randeria, M. Precursor pairing correlations and pseudogaps, in Proc. Int. School of Physics 'Enrico Fermi' on Conventional and High Temperature Superconductors (ed. Iadonisi, G., Schrieffer J. R., Chiafalo, M.L.) 53-75 (IOS Press, Amsterdam 1998).

16. Vershinin, M. *et al.* Local ordering in the pseudogap state of the high-$T_c$ superconductor $Bi_2Sr_2CaCu_2O_{8+\delta}$. *Science* **303**, 1995-1998 (2004).

17. Hanaguri, T. *et al.* A 'checkerboard' electronic crystal state in lightly hole-doped $Ca_{2-x}Na_xCuO_2Cl_2$. *Nature* **430**, 1001-1005 (2004).

18.    McElroy, K. *et al.* Coincidence of checkerboard charge order and antinodal state decoherence in strongly underdoped superconducting $Bi_2Sr_2CaCu_2O_{8+\delta}$. *Phys. Rev. Lett.* **94**, 197005 (2005).

19.    Renner, Ch., Revaz, B., Genoud, J.-Y., Kadowaki, K. & Fischer, Ø. Pseudogap precursor of the superconducting gap in under- and overdoped $Bi_2Sr_2CaCu_2O_{8+\delta}$. *Phys. Rev. Lett.* **80**, 149-152 (1998).

20.    Kugler, M., Fischer, Ø., Renner, Ch., Ono, S. & Ando, Y. Scanning tunneling spectroscopy of $Bi_2Sr_2CuO_{6+\delta}$: New evidence for the common origin of the pseudogap and superconductivity. *Phys. Rev. Lett.* **86**, 4911-4914 (2001).

21.    Deutscher, G. Coherence and single particle excitations in high temperature superconductors. *Nature* **397**, 410-412 (1999).

22.    Andersen, B. M., Melikyan, A., Nunner, T. S. & Hirschfeld, P. J. Thermodynamic transitions in inhomogeneous d-wave superconductors. *Phys. Rev. B* **74**, 060501(R) (2006).

23.    Norman, M. R. *et al.* Destruction of the Fermi surface in underdoped high-$T_c$ superconductors. *Nature* **397**, 157-160 (1998).

24.    Valla, T., Fedorov, A. V., Lee, J., Davis, J. C. & Gu, G. D. The ground state of the pseudogap in cuprate superconductors. *Science* **314**, 1914-1916 (2006).

25.    Tanaka, K. *et al.* Distinct Fermi-momentum-dependent energy gaps in deeply underdoped Bi2212. *Science* **314**, 1910-1913 (2006).

26.    Le Tacon, M. *et al.* Two energy scales and two distinct quasiparticle dynamics in the superconducting state of underdoped cuprates. *Nature Physics* **2**, 537-543 (2006).

27.    Damascelli, A., Hussain, Z. & Shen, Z.-X. Angled-resolved photoemission studies of cuprate superconductors. *Rev. Mod. Phys.* **75**, 473-541 (2003).

28.    Carbotte, J. P. Properties of boson-exchange superconductors. *Rev. Mod. Phys.* **62**, 1027-1157 (1990).

29.    Balatsky, A. V. & Zhu, J.-X. Local strong coupling pairing in d-wave supercondcutors with inhomogeneous bosonic modes. *Phys. Rev. B* **74**, 094517 (2006).

**Supplementary Information** is linked to the online version of the paper at www.nature.com/nature.

**Acknowledgements** This work was supported by Princeton University, the NSF-DMR, and the NSF-MRSEC programme through the Princeton Center for Complex Materials.

**Author Information** Reprints and permissions information is available at www.nature.com/reprints. The authors declare no competing financial interests. Correspondence and requests for materials should be addressed to A.Y. (yazdani@princeton.edu).

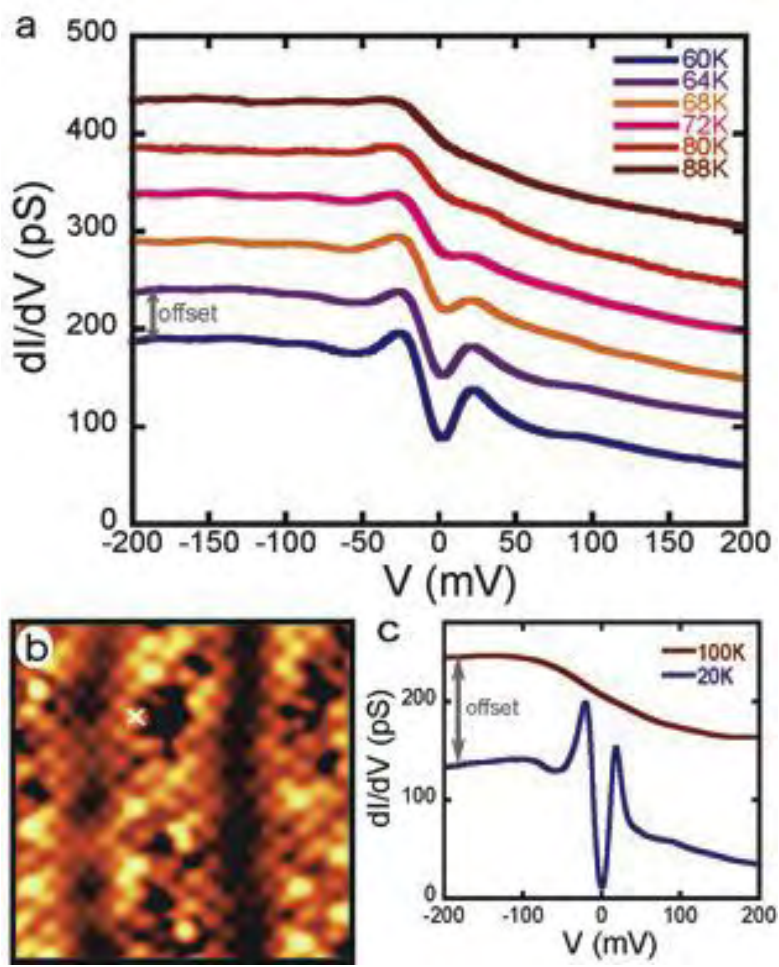

**Figure 1 | Spectroscopy at a specific atomic site as a function of temperature (d*I*/d*V* versus sample bias *V*).** Samples were cleaved *in situ* at room temperature before being inserted into the cooled microscope stage and measured at various temperatures (20–180 K). **a,** Spectra taken at different temperatures at the same location in an OV65 ($T_c$ = 65 K) sample (offset for clarity). Special experimental procedures have been used to track the same location on the sample within 0.1 Å at different temperatures. **b,** Topography of the sample showing the location for spectra in **a**. Between each measurement the STM is stabilized at each temperature for a period of up to 24 h, during which STM topography, such as that shown in **b**, is used to centre the location of spectroscopic measurements shown in **a**. **c,** Typical spectra at 20 and 100 K for the same sample (not the same location). The coherence peaks become more pronounced at lower temperatures, but the position of the peak does not change significantly below 60 K. Analysis of the low-temperature spectra (see Supplementary Information) shows them to be consistent with those of the excitation spectrum of a *d*-wave superconductor uniformly averaged around the Fermi surface with a maximum pairing gap corresponding to $\Delta \approx 24$ mV.

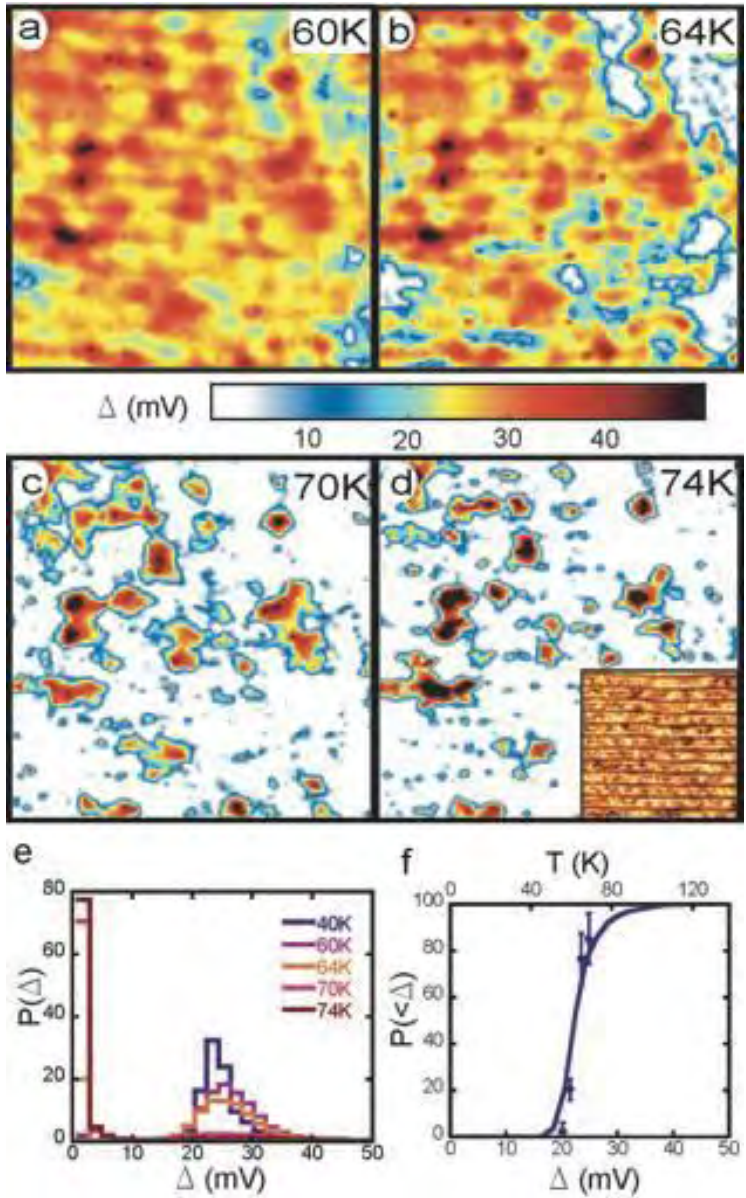

**Figure 2 | Gap evolution for overdoped ($T_c = 65$ K) samples.** **a–d**, Gap maps taken on the same 300 Å area of an OV65 sample at different temperatures near $T_c$. At each temperature and atomic site, the value of the gap can be extracted from local spectroscopic measurements by using the experimental criterion that the local d$I$/d$V$ has a maximum at $V = +\Delta$, as demonstrated by data in Fig. 1a. The gaps vary spatially on the scale of 1–3 nm. The inset to **d** shows the topography of the area. **e**, The probability of finding a gap of a given size (gap distribution) for the gap maps shown in (**a–d**) and an additional gap map taken at 40 K. **f**, The solid line shows the probability P($<\Delta$) that the gaps are less than a given $\Delta$ (lower x axis). This is obtained at a given voltage by summing the lowest temperature histogram of $\Delta$ to that voltage. The percentage of ungapped area of the sample (points) is plotted as function of temperature (shown on the upper x axis). The scaling between the two x axes is $2\Delta/k_B T = 7.8$. The $1\sigma$ error bars were calculated from the finite statistics (arising from the limited area of each map) and the energy and conductance resolution of the spectra used to compose the maps.

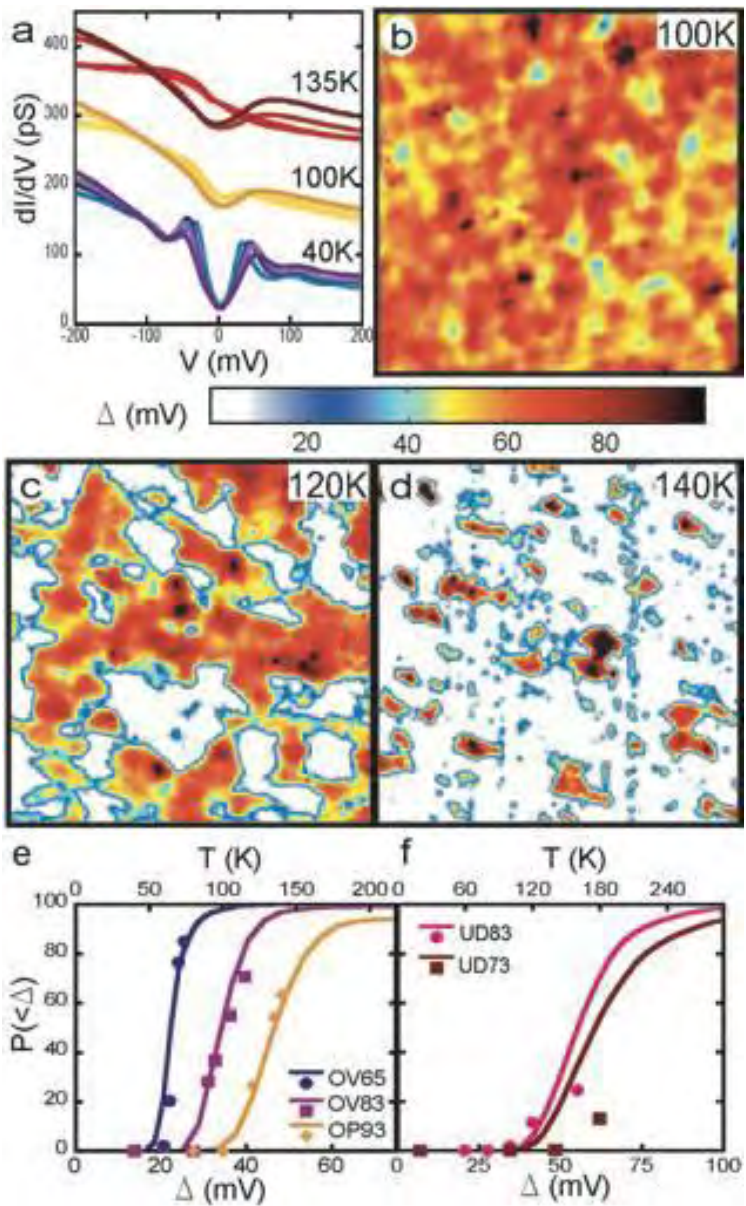

**Figure 3 | Gap evolution at different dopings. a**, Representative spectra taken at different temperatures (offset for clarity) on an OP93 ($T_c = 93$ K) sample. The coherence peaks diminish in intensity on raising the temperature through $T_c$, but a peak in the spectrum persists at positive bias. **b–d**, Gap maps taken over 300 Å areas at different temperatures on the OP93 sample. Although the area over which the maps have been obtained is different for each temperature, the gap distributions are statistically meaningful because of the large areas over which data has been collected. **e**, The solid lines show the integral of the gap distributions versus gap size (lower *x* axis) for three different doping levels, from overdoped ($T_c = 65$ and 83 K) to optimally doped ($T_c = 93$ K). The points show the percentage of the ungapped area of the samples as a function of temperature (upper *x* axis). The scaling between the two axes is $2\Delta/k_BT = 8.0$. **f**, The solid lines show the integral of the gap distributions versus gap size (lower axis) for two different doping levels, both underdoped ($T_c = 83$ and 73 K). The points show the percentage of ungapped area of the samples as a function of temperature (upper *x* axis). The scaling between the two axes is $2\Delta/k_BT = 8.0$.

**Figure 4 | Pseudogap and pairing gaps in underdoped samples. a**, Representative spectra taken at specific locations well above $T_c$ for all doping levels (offset for clarity; bar marks zero conductance for offset spectra. **b**, Spectra taken at 20 K at various locations on the UD73 sample. Within the large gap we see a 'kink' at lower energy, indicated by the arrow. Inset, distributions of the large gap and the 'kink' energy on the UD73 sample at 20 K.

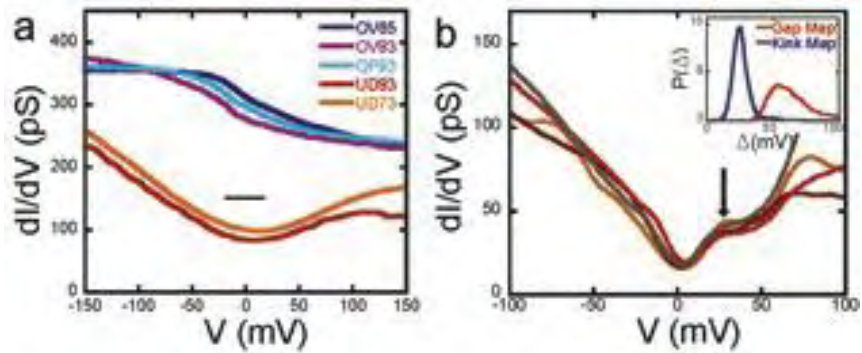

**Figure 5 | Schematic phase diagram for Bi₂Sr₂CaCu₂O₈₊δ.** Temperatures and doping levels where large area gap maps were obtained are indicated by points. The colours are the percentage of the sample that is gapped at a given temperature and doping as measured in the gap maps. The $T_{p, max}$ line is the temperature at which <10% of the sample is gapped. The lower solid line denotes the bulk $T_c$.

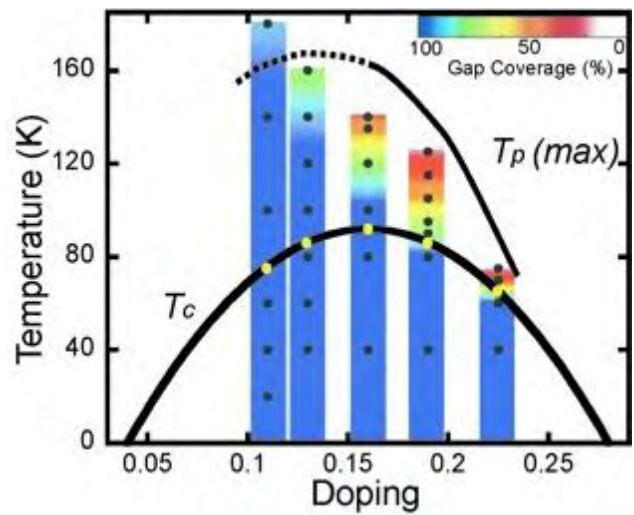

# SUPPLEMENTARY INFORMATION

# Visualizing Pair Formation on the Atomic Scale in the High-$T_c$ Superconductor Bi$_2$Sr$_2$CaCu$_2$O$_{8+\delta}$

Kenjiro K. Gomes[1,*], Abhay N. Pasupathy[1,*], Aakash Pushp[1,*], Shimpei Ono[2], Yoichi Ando[2] & Ali Yazdani[1]

## 1. D-wave fits to the low temperature gap

At temperatures well below the superconducting transition temperature $T_c$, we see a single gap in the spectrum in optimally doped and overdoped samples, accompanied by a bias-asymmetric background (Fig. 1a). The spectral line shape within the gap can be fit with the d-wave pairing DOS averaged equally around all angles. This d-wave density of states is

$$\rho(E,T) = \frac{1}{\pi} \int_0^\pi d\theta \, Re \left[ \frac{|E' - i\Gamma(T)|}{\sqrt{(E' - i\Gamma(T))^2 - \Delta(T)^2 \cos^2(2\theta)}} \right] \tag{1}$$

where $\Gamma$ is an energy independent lifetime broadening[30]. At low temperatures we set $\Gamma=0$[30]. The differential conductance can be fit by the form

$$\frac{dI}{dV} = a \int_{-\infty}^{\infty} dE \frac{df(E+V)}{dV} \rho(E,T) + bV \tag{2}$$

Here $\rho(E,T)$ is the d-wave superconductor DOS and $f(E)$ is the Fermi distribution. The scale parameter $a$ is adjusted to match the measured conductance. We fit the normal state background with a line within the gap using the parameter $b$. From the shape of the normal state (Fig. 1a,c), we find that such an assumption for the background works well for gaps <60 mV.

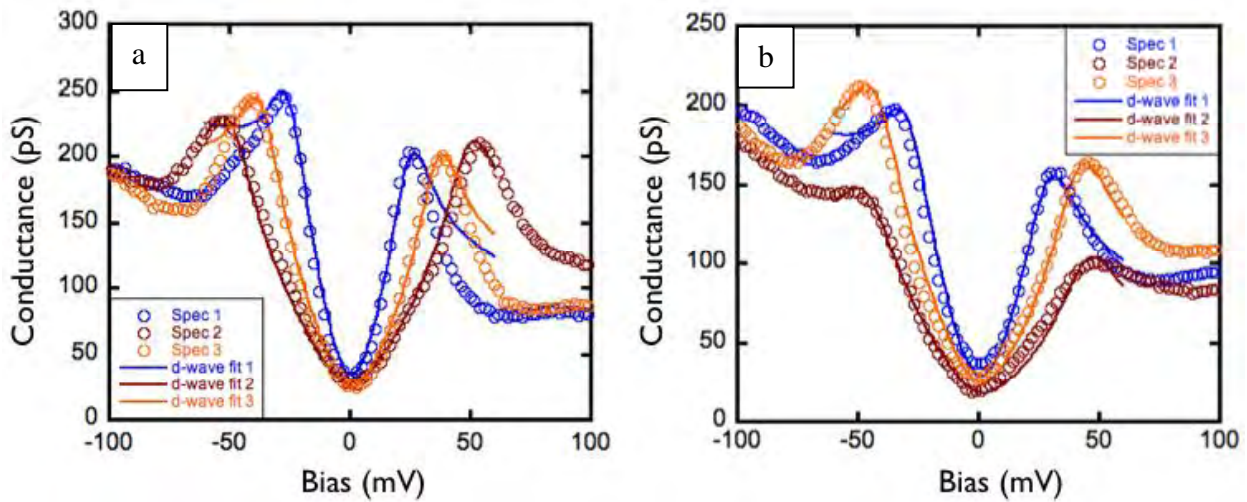

**Figure S1**: Low temperature d-wave fits to the gap. **a**. d-wave fits to the OV65 sample at 40 K. A single d-wave gap is used to fit the spectrum at each point, with a linear background. Fits are shown for three gaps of different sizes taken at different spatial locations. The fits work well within the gap region but deviate from the data beyond the gap[30]. **b**. Similar fits for the OP93 sample at 40 K.

Sample fits to this DOS shape are shown for the overdoped ($T_c = 65$K) sample (Fig. S1a) and for the optimally doped sample (Fig. S1b) for gaps of different sizes. In general, all gaps smaller than 60 mV can be fit well with this procedure within the gap region. Beyond the gap region, effects such as strong coupling to bosonic modes[31] cause deviations from the d-wave fits. The extracted gap value is 5-10% smaller than the position of the conductance peak. Given that the peak position does not vary much with temperature, we expect this agreement between the fit value and the peak position to only improve at lower temperature. Importantly, such fits show that the low temperature spectra can be described by a simple d-wave pairing gap without any k-dependent matrix elements. This result is different from tunnel junctions with a large overlap area, where a k-dependent matrix element is needed to achieve good fits to the data[30]. We have also checked that the extracted gap is independent of the height of the tip (for junction resistances > 1 GΩ), indicating that our tip is well approximated by a point probe.

## 2. Evolution of gap distributions across $T_c$

As the temperature is raised through $T_c$, spectra at individual locations evolve smoothly without any discontinuous changes at $T_c$. In the most overdoped sample studied (OV65), a significant percentage of the sample (~ 20 %) loses its gap at the bulk transition temperature as shown in Fig. 2. This however is no longer the case as we move towards optimal doping. For the overdoped sample with $T_c =83$ K, a significant portion of the sample is gapped only at 95 K. At optimal doping, the entire sample remains gapped as we go through $T_c$. The distribution of gaps at 100 K ($T_c = 93$ K) is comparable to the distribution of gaps seen at 80 K, as shown in Fig. S2. The difference seen in the two histograms implies that the spectral peak position might shift on raising the temperature. We believe that such a shift can be accounted for by thermal broadening via two mechanisms. Firstly, the Fermi distribution becomes broader with increasing temperature due to which the conductance peak moves to higher voltage[32]. Secondly, the quasiparticle lifetime broadening is known to increase as T goes through $T_c$[33]. Both these effects can be modeled using the d-wave DOS in eqn. (1). We find that a 40 mV gap (typical of optimally doped samples) at T=0 has a peak position that increases with temperature. The shift due to the Fermi distribution is ~ 3mV between 80 K and 100 K. This, together with the addition of a modest lifetime broadening (5-10 meV) can explain the shift in the peak position between 80 K and 100 K. The difference in the sharpness of histograms between 40 K and 80 K is comparable to the sample-to-sample variation. The sample-to-sample variations have little impact on the calculation of the local criterion $2\Delta/k_B T_p$, generating deviations within our error bars.

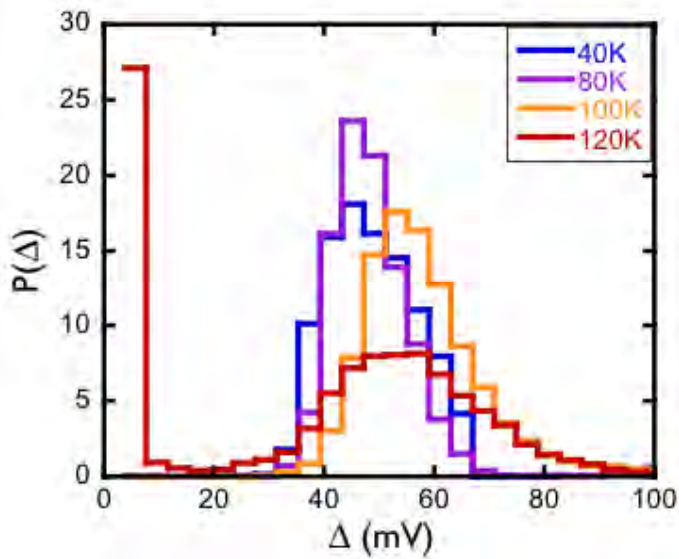

**Figure S2**: Gap distributions for optimally doped samples taken at 40 K, 80 K (superconducting), 100 and 120 K (non-superconducting). The gaps evolve continuously through $T_C$ (93 K) and all the gaps survive at 100 K. Gaps are first destroyed at 110-120 K.

The significant feature of Fig. S2 which we wish to highlight is that a gap exists at every point in space at 100 K in the optimally doped sample. To make a significant proportion of the sample lose its gap, the temperature has to be raised to ~ 25 K above $T_c$.

### 3. Scaling of gap distributions for different doping levels.

The mean gap is a decreasing function of doping as seen in Table 1. We find that the width of any given gap distribution at low temperature scales with its mean value[30]. Shown in Fig. S3 are histograms of the gap distributions for the five samples discussed.

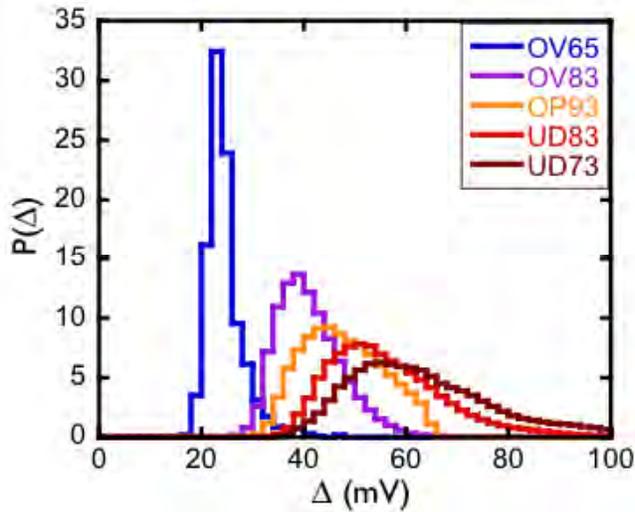

**Figure S3**: Gap map distributions taken in the superconducting state for the five doping levels studied. The distributions are recorded at 40 K for sample OV65, 40 K for sample OV83, 40 K for sample OPT, 50 K for sample UD83 and 20 K for sample UD73.

### 4. Representative spectra for the OV65 sample.

In all samples studied the gap varies in space. Shown in Fig. S4a are spectra taken at 40 K along a 130 Å line in the OV65 sample showing the typical variation of the spectra from point to point. Well below $T_c$, the spectra all show a gap and are homogeneous at low bias. This is not true above $T_c$, where there is significant inhomogeneity at low bias (Fig. S4b), with parts of the sample losing their gap.

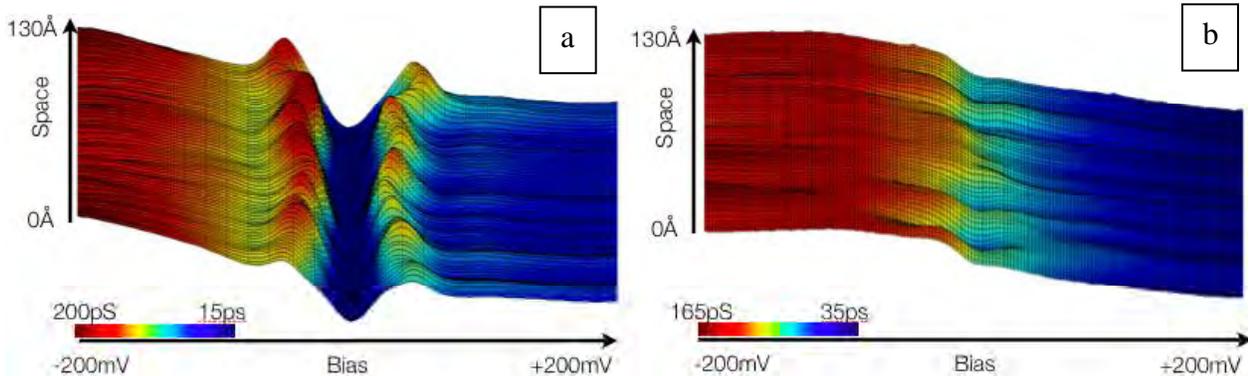

**Figure S4**: Representative spectra taken on the OV65 sample. **a**. Spectra taken along a 130Å line at 40 K in the superconducting state. **b**. Spectra taken at 70 K (Tc=65 K).

## 5. Representative spectra for the OPT sample.

The optimally doped sample (Tc=93 K) shows well-developed coherence peaks at 40 K. Shown in fig S5a is a series of spectra taken along a 130 Angstrom line. At 100 K, the entire sample still remains gapped, with a spatially inhomogeneous gap as shown in fig S5b. On raising the temperature to 135 K, a significant proportion of the sample has lost its gap (fig S5c).

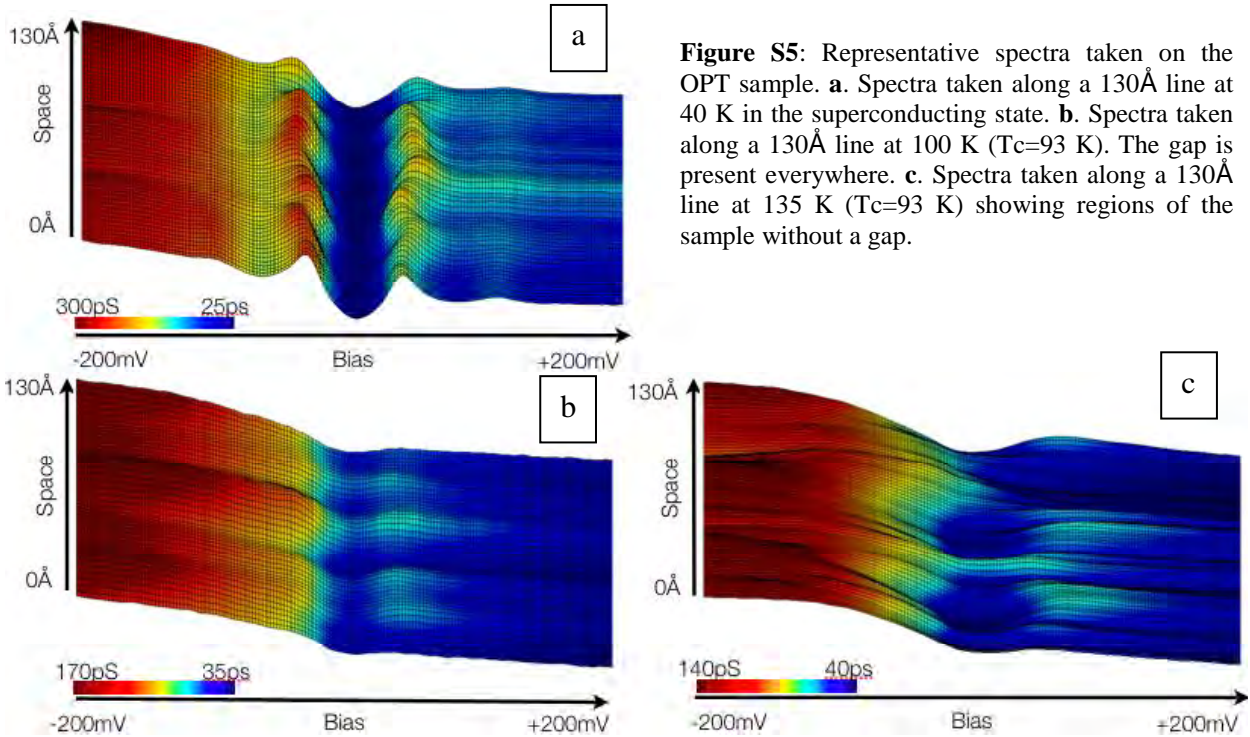

**Figure S5**: Representative spectra taken on the OPT sample. **a**. Spectra taken along a 130Å line at 40 K in the superconducting state. **b**. Spectra taken along a 130Å line at 100 K (Tc=93 K). The gap is present everywhere. **c**. Spectra taken along a 130Å line at 135 K (Tc=93 K) showing regions of the sample without a gap.

## 6. Representative spectra for the UD73 sample.

The heavily underdoped sample (UD73) shows multiple gap features at low temperature as shown in Fig. S6a. These features are seen in ~30% of the sample at 20 K. On raising the sample temperature to well above $T_c$, the smaller gap features disappear and a single large gap is seen in the spectrum (fig S6b).

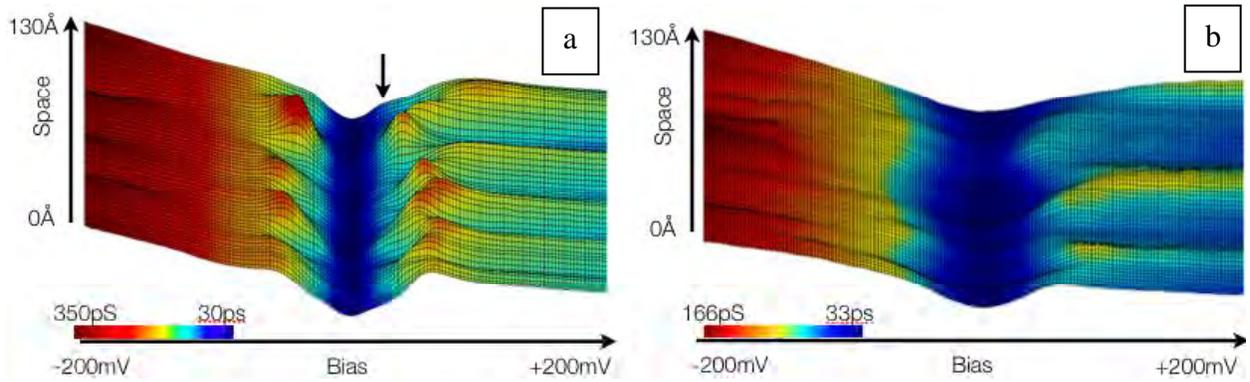

**Figure S6**: Representative spectra taken on the UD73 sample. **a.** Spectra taken along a line 130Å at 20 K in the superconducting state. A second gap feature is seen at low bias in ~30% of the sample **b.** Spectra taken along a 130Å line at 80 K (Tc=73 K).

## 7. Statistics of samples studied

Shown in table 1 are statistical properties of the five doping levels discussed in the article. The sample $T_c$ is found by susceptibility measurements performed with a SQUID. The average gap ($\Delta_{avg}$) is extracted from the gap maps at the lowest temperatures studied. The ratio $2\Delta/k_B T_p$ is extracted from the best fit between the low temperature gap histograms and the gap maps at various temperatures. In Fig. S7, we illustrate the appropriateness of the local criterion $2\Delta/k_B T_p = 8$, in contrast to other ratios. We replot the same data shown in Fig. 3e using different values for the local ratio to illustrate how they poorly represent our data.

**Table 1 | Samples and Gap Map Statistics**

| Sample code | $T_c$ (K) | Doping | $\Delta_{avg}$ (mV) | $2\Delta/k_B T_p$ |
|:---:|:---:|:---:|:---:|:---:|
| OV65 | 65 | 0.22 | 26 | 7.8 ± 0.3 |
| OV83 | 88 | 0.19 | 38 | 7.8 ± 0.2 |
| OP93 | 93 | 0.16 | 52 | 8.1 ± 0.2 |
| UD83 | 83 | 0.14 | 61 | 7.9 ± 0.6 |
| UD73 | 73 | 0.12 | 66 | * |

* Cannot extract from data.

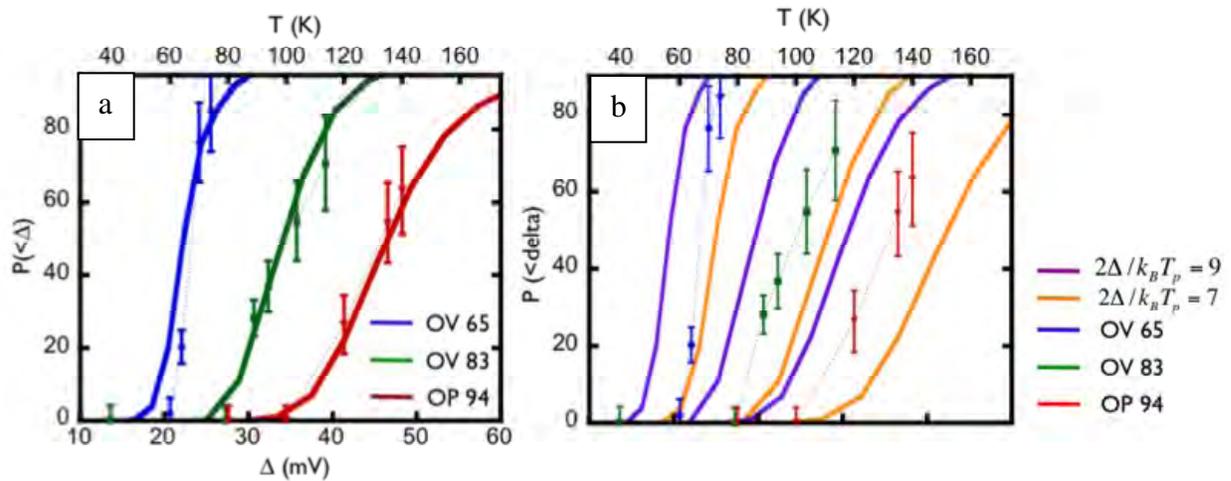

**Figure S7:** The same data as the plot shown in Fig. 3e of the main paper. The points represent the percentage of ungapped regions in gap maps as function of temperature. The solid lines are the integral of the low temperature gap histograms. **a.** The ratio between temperature and energy scales is given by $2\Delta/k_B T_p = 8$. **b.** Same data used in **a** but using different scaling ratios to illustrate the effect of deviations from the ratio $2\Delta/k_B T_p = 8$ used in the paper. The purple solid lines are the gap histogram integrals plotted using $2\Delta/k_B T_p = 9$ and the orange lines are using $2\Delta/k_B T_p = 7$.

## References

[30]    L. Ozyuzer, J. F. Zasadzinski, K. E. Gray, C. Kendziora and N. Miyakawa, Absence of pseudogap in heavily overdoped $Bi_2Sr_2CaCu_2O_{8+\delta}$ from tunneling spectroscopy of break junctions, *Europhys. Lett.* **58**, 589 (2002).

[31]    J. Lee et al. Interplay of Electron-Lattice Interactions and Superconductivity in $Bi_2Sr_2CaCu_2O_{8+\delta}$ , *Nature* **442**, 546-550 (2006).

[32]    I. Giaver and K. Megerle, Study of Superconductors by Electron Tunneling, *Phys. Rev.* **122**, 1101 (1961).

[33]    M.R. Norman, M. Randeria, H. Ding and J.C. Campuzano, Phenomenology of the low-energy spectral function in high-Tc superconductors, *Phys. Rev. B* **57**, R11093 (1998) .